 \documentstyle[12pt]{article}
 
\begin{document}

\title{On  Expansion in the Width for Domain Walls
\thanks{Paper supported
in part by the grant KBN 2 P302 049 05.}}

\author{by\\
\\
H. Arod\'z   \\
\\
Institute of Physics, Jagellonian University,
Cracow \thanks{Address: Reymonta 4, 30-059 Cracow, Poland.}
        \thanks{E-mail: ufarodz@ztc386a.if.uj.edu.pl} }

\date{  $\;\;$}
\maketitle

\thispagestyle{empty}
 $\;\;\;$ \\
{\bf Abstract} \\
The well-known idea to construct  domain wall type solutions of
field equations by means of an expansion in the width of the domain
wall is reexamined. We observe that the problem involves singular
perturbations. Hilbert-Chapman-Enskog method is used to construct
a consistent perturbative expansion. We obtain the solutions to
the second order in the width without introducing an effective action
for the domain wall. We find that zeros of the scalar field in general
do not lie on a Nambu-Goto trajectory. As examples we consider cylindrical
and spherical domain walls. We find that the spherical domain wall,
in contradistinction to the cylindrical one, shows an effective rigidity.
$\;\;$ \\
January 1995     \nopagebreak    \\
TPJU-2/95

\pagebreak

\setcounter{page}{1}
\section{Introduction}

Dynamics of topological defects is to be deduced from
underlying nonlinear equations describing evolution of fields the defect is
composed of. In general, it is a formidable task and one is forced to
resort to approximate analytical solutions or to numerical solutions. The
results are of great interest for particle physics (e.g. for dynamics of
a flux-tube in QCD \cite{1}), for field-theoretical
cosmology (e.g. cosmic strings \cite{2}), and for condensed matter
physics (e.g. domain walls in magnetics \cite{3}, vortices in
superconductors \cite{4} or in superliquid Helium \cite{5}, defects in
liquid crystals \cite{6}).

In particular, in the context of relativistic field theoretical models there
is a long-standing, interesting and difficult problem how to calculate
evolution of vortices and of domain walls. Theoretical descriptions of
vortices and domain walls show many similarities, and the two kinds of
objects can be treated by essentially identical methods. The problem has
become even more interesting  after the seminal paper \cite{7} has appeared,
in which it has been  pointed out that the vortex dynamics can
approximately be described in terms of a string model. Apart from
purely numerical computations, see e.g. \cite{8,9,10}, the predominant
approach is the effective action method, started in the paper \cite{11}
and pursued in numerous papers till nowadays, see e.g.
 \cite{12,13,14,15,16,17,18,19}.
The idea is to reduce the complicated field-theoretical dynamics to much
simpler dynamics of a string or a membrane by regarding  some degrees of
 freedom as inessential. In the first step, these degrees of freedom are
 eliminated by  the assumption that  they have definite
  values determined from  a conditional stationary action principle
for the original  field-theoretical action. The condition usually adopted is
that zeros of the scalar field have apriori fixed, arbitrary trajectory
in Minkowski space-time (identified with the world-volume of the
membrane). Inserting these conditional
solutions into the original action one obtains the effective action for
the membrane. In the second step one has to find solutions of the equations
of motion for the membrane, derived from the  action principle for
the effective action. Let us stress that in this approach
calculation of the effective action is the necessary intermediate
step. Without it, one has only the whole set of conditional
 solutions which exist for all trajectories of the membrane, including
 unphysical ones. It is just the role of the second step to pick physical
 trajectories for the membrane.     See the paper \cite{15} for an
 explicit formulation and \cite{20} for a discussion of this method.

Another group of approaches aims at deriving directly from the original
field equations dynamical equations  for certain characteristics of the
topological defect, e.g. for an average radius in the case of a ring-like
vortex \cite{21}, or position of zeros of the Higgs field \cite{22}. In these
approaches one does not have to calculate the effective action as the
 necessary intermediate step.
To this group belongs also the approach based on a polynomial approximation
to the fields constituting the defect \cite{25} -- in this approach also
 a width of the domain wall is  a dynamical variable.

In the present paper we would like to discuss one more method of calculating
the evolution  of domain walls or vortices directly from the original field
equations: an expansion in the width of the domain wall or a vortex. This
method has been tried in many papers, e.g. in \cite{11}-\cite{19},
 always in conjunction with the effective action
  method. In the present paper we show
 that the expansion in the width itself, when appropriately carried out, is
 sufficient to determine evolution of the defect. No effective action is
 necessary.  We will describe our formulation of this method in the case
 of the domain walls. Analogous calculation for vortices will be presented
 as a separate paper.

Our main steps  are moreless the same as in the previous papers:
we introduce a special coordinate system co-moving with
the domain wall and we formally expand the sought for solution of the field
equation in the positive powers of the width of the domain wall.
(The actual expansion parameters are dimensionless ratios of the width
 with  curvature radia of the domain wall.) The first new feature specific
 for our formulation is that we do not tie the co-moving coordinate system
 with position of the zeros of the  scalar field. It is tied to a
 Nambu-Goto  membrane --  this fact  follows from the field
 equation as a   consistency condition. However this
 membrane in general is not the locus of the zeros of the field.
  The position of the   zeros is  determined afterwards, from the
 explicitely constructed  approximate solution of the field equation.

 Second, we observe that the expansion in the width is an example of
 so called singular perturbations \cite{24}. In such case one has to
 take into account certain consistency conditions for the  perturbative
 expansion.  Using Hilbert-Chapman-Enskog method  for constructing
 perturbative expansion in the case of singular perturbations,
see \cite{25}  for a lucid description of it,
 we obtain the domain wall solution  directly by approximate solving
  the original  field equation, without any need to introduce
 the effective action.

 The plan of our paper is as follows. In Section 2 we introduce the co-moving
 coordinates and we present the field equation. This section is
 included mainly for convenience of the reader. In Section 3 we apply the
Hilbert-Chapman-Enskog method to calculate  the domain wall solution in the
 perturbative expansion in positive powers of the width. In Section 4 we
present simple examples: solutions  describing cylindrical and
  spherical domain walls.   Section 5  contains ending remarks.

\section{Transformation to the co-moving coordinates}

We shall consider domain walls in a well-known model defined by the
 following Lagrangian
\begin{equation}
 L=-\frac{1}{2}\eta_{\mu\nu}\partial^{\mu}\Phi \partial^{\nu}\Phi
-\frac{\lambda}{2}(\Phi^{2}-\frac{M^2}{4\lambda})^2,   \end{equation}
where $(\eta_{\mu\nu})=diag(-1,1,1,1)$, and $\lambda,M$ are positive
constants. The corresponding field equation for the scalar field $\Phi$
has the  form
\begin{equation} \partial_{\mu}\partial^{\mu}\Phi -
2 \lambda (\Phi^2-\frac{M^2}{4\lambda})\Phi=0. \end{equation}

There are two vacuum values of the field $\Phi$  equal to $\pm \Phi_0$,
where $\Phi_0 \equiv M/2\sqrt{\lambda}$. The domain wall is given by the
field $\Phi$ smoothly interpolating between the two vacua.
At each instant of time the field $\Phi$
vanishes somewhere  inside the domain wall. We assume that the locus of
these zeros is a smooth surface $S$.
 We shall call it the core of the domain wall. Well-known example
  of the domain wall, with the core given by the
$(x^1,x^2)$ plane, is given by the following exact, static solution of Eq.(2)
\begin{equation}
\Phi = \Phi_0\; tanh(\frac{x^3}{2l_0}),
\end{equation}
where $l_{0} \equiv M^{-1}.$  The width of this domain wall is of the order
$l_{0}$, and energy density is exponentially localised around the $(x^1,x^2)$
plane.

For a  generic domain wall, we parametrise the  world-volume $\Sigma$
of the core (a 3-dimensional manifold embedded in Minkowski space-time,
whose time slices coincide with $S$) as follows
\begin{equation}
\left( X^{\mu}\right) (\tau,\sigma^1,\sigma^2) = \left(  \begin{array}{c}
\tau \\  X^1(\tau,\sigma^1,\sigma^2) \\ X^2(\tau,\sigma^1,\sigma^2) \\
X^3(\tau,\sigma^1,\sigma^2)
\end{array} \right),
\end{equation}
where $\tau$ coincides with the laboratory frame time $x^{0}$, and $\sigma^1,
\sigma^2$ parametrise the core $S$ at each  instant of time.

We shall also need another surface $\tilde{S}$ attached to the domain wall. In
general $\tilde{S} \neq S$. In the next section we shall obtain a Nambu-Goto
equation for $\tilde{S}$, so this surface can be regarded as the Nambu-Goto
type relativistic membrane. The surface $\tilde{S}$ is related to the core
$S$ by a simple formula (51) derived in Section 4. In the following we shall
call $\tilde{S}$ the membrane co-moving with the domain wall. Let us
stress that the membrane is merely an auxilliary mathematical
 constructs -- the physical object is the domain wall.

The world-volume of the membrane will be denoted by $\tilde{\Sigma}$, and we
shall parametrise it as follows
\begin{equation}
\left( Y^{\mu}\right) (\tau,\sigma^1,\sigma^2) = \left(  \begin{array}{c}
\tau \\  Y^1(\tau,\sigma^1,\sigma^2) \\ Y^2(\tau,\sigma^1,\sigma^2) \\
Y^3(\tau,\sigma^1,\sigma^2)
\end{array} \right),
\end{equation}
where $\tau$ coincides with the laboratory frame time $x^{0}$. The
parameters $\sigma^1,\sigma^2$ in formula (5) are in principle
 independent of the ones in formula (4). We use the same notation for
  convenience, and also in anticipation of a one-to-one correspondence
 between $S$ and $\tilde{S}$ given by formula (51).

As usual, we introduce a special coordinate system
$(\tau,\sigma^1,\sigma^2,\xi)$ co-moving with the domain wall.
  These coordinates  are defined by the following formula
\begin{equation}
x^{\mu} = Y^{\mu}(\tau,\sigma^1,\sigma^2) + \xi \: n^{\mu}(\tau,\sigma^1,
\sigma^2),
\end{equation}
where $x^{\mu}$ are Cartesian, laboratory frame coordinates in Minkowski
space-time, and $(n^{\mu})$ is a normalised space-like four-vector,
 orthogonal to the $\tilde{\Sigma}$ in the covariant sense, i.e.
\begin{equation}
n_{\mu}(\tau,\sigma^1,\sigma^2)Y^{\mu}_{,a}(\tau,\sigma^1,\sigma^2) = 0,
  \;\;\;\;\; n_{\mu}n^{\mu}=1,
\end{equation}
where $a=0$ corresponds to $\tau$; $a=1, a=2$ correspond
to $\sigma^1,\sigma^2$; $Y^{\mu}_{,\tau}\equiv \partial Y^{\mu}/
\partial\tau$, etc.  The four-vectors $Y_{,\tau}, Y_{,\sigma^1},
 Y_{,\sigma^2}$ are tangent to $\tilde{\Sigma}$. For points lying on the
 membrane $\xi = 0$, and the parameter $\tau$ coincides with the
 lab-frame time $x^{0}$.
For $\xi \neq 0$,  $\tau$ in general is not equal to the lab-frame time $x^0$.
Notice that definition (6) implies that $\xi$ is a Lorentz scalar.
In the co-moving coordinates the membrane is described by the simple,
Lorentz invariant condition $\xi=0$.

The next step is to write Eq.(2) in the new coordinates. It is convenient to
 introduce extrinsic curvature coefficients $K_{ab}$ and induced
metrics $g_{ab}$ on $\tilde{\Sigma}$:
\begin{equation}
K_{ab} = n_{\mu} Y^{\mu}_{,ab}, \;\;\;\;  g_{ab}=Y^{\mu}_{,a} Y_{\mu,b},
\end{equation}
where $a,b=0,1,2$. The covariant metric tensor in the new coordinates
has the following form
\begin{equation}
[G_{\alpha\beta}] = \left[ \begin{array}{lr} G_{ab} & 0 \\
0 & 1 \end{array} \right],
\end{equation}
where $\alpha,\beta=0,1,2,3$;\  $\;\;\;\alpha=3$ corresponds to the $\xi$
coordinate; and
\begin{equation}
G_{ab}=N_{ac} g^{cd} N_{db},\;\;\; N_{ac}\equiv g_{ac} - \xi K_{ac}.
\end{equation}
Thus, $G_{\xi\xi}=1, \; G_{\xi a}=0$ (a=0,1,2, as in (7)).
It follows from formula (10) that
\begin{equation}
\sqrt{-G} = \sqrt{-g} \; h(\tau,\sigma^1,\sigma^2,\xi),
\end{equation}
where as usual  $g\equiv det[g_{ab}]$, $G\equiv det[G_{\alpha\beta} ]$, and
\begin{equation}
h(\tau,\sigma^1,\sigma^2,\xi) =
1 - \xi K^{a}_{a} +\frac{1}{2} \xi^2 (K^{a}_{a} K^{b}_{b} - K^{b}_{a}
K^{a}_{b}) -\frac{1}{3!} \xi^3
\epsilon_{abc}\epsilon^{def} K^{a}_{d} K^{b}_{e} K^{c}_{f}.  \nonumber
\end{equation}
Also $g$ can depend on $\tau,\sigma^1,\sigma^2$.
For raising and lowering the latin indices of the extrinsic
curvature coefficients we use the induced metric tensors
$g^{ab},\;\; g_{ab}$.

The inverse metric tensor $G^{\alpha \beta}$ is given by
\begin{equation}
[G^{\alpha \beta}] = \left[ \begin{array}{lr} G^{ab} &  0 \\
0 &   1 \end{array} \right],
\end{equation}
where
\begin{equation}
G^{ab}= (N^{-1})^{ac} g_{cd} (N^{-1})^{db}.
\end{equation}
Explicit formula for $(N^{-1})^{ac}$ has the following form:
\begin{equation}
(N^{-1})^{ac} = \frac{1}{h} \left\{ g^{ac} [1- \xi K^{b}_{b} +\frac{1}{2}
\xi^2 ( K^{b}_{b} K^{d}_{d} - K^{d}_{b} K^{b}_{d} )] \right.
\end{equation}
\[ \;\;\;\;\; +   \left. \xi (1-\xi K^{b}_{b}) K^{ac} + \xi^2
 K^{a}_{d} K^{dc} \right\}  \]
(this is just the matrix inverse to $[N_{ab}]$; by definition it has
upper indices).

 In general, the coordinates $(\tau, \sigma^1, \sigma^2, \xi)$ are
 defined locally,  in a vicinity of the world-volume $\tilde{\Sigma}$.
Roughly speaking, the allowed range of the $\xi$ coordinate is restricted
by the smallest of the two curvature radia of the membrane in the local
rest frame of a piece of the membrane.
 We assume that the domain wall is sufficiently narrow so that outside of
 the region of validity of the co-moving coordinates there are only
 exponential tails of the domain wall, i.e. the solution is
 exponentially close to one of the two vacuum solutions.
 Detailed discussion of the region of validity of the
co-moving coordinates has been given in \cite{23}.

 In the co-moving coordinates the field equation (2) has the following form
 \begin{equation}
  \frac{1}{\sqrt{-G}}\frac{\partial}{\partial u^{a}}
(\sqrt{-G}\;G^{ab}
\frac{\partial \Phi}{\partial u^{b}})
  + \frac{1}{h} \partial_{\xi}( h \;\partial_{\xi}\Phi)
  -2 \lambda  (\Phi^2 -\frac{M^2}{4\lambda} ) \Phi =0,
 \end{equation}
where $u^0 \equiv \tau, u^1\equiv \sigma^1, u^2 \equiv \sigma^2$.

Let us rescale the scalar field and the $\xi$ coordinate
\begin{equation}
\Phi(x^{\mu}) \equiv \frac{M}{2\sqrt{\lambda}} \phi(u^a,s),
 \;\; \xi \equiv \frac{2}{M} s,
\end{equation}
where $\phi(u^a,s)$ and $s$ are dimensionless. Then, equation (16)
has the form
 \begin{equation}
  \frac{2}{M^2}\frac{1}{\sqrt{-G}}\frac{\partial}{\partial u^{a}}
(\sqrt{-G}\;G^{ab}
\frac{\partial \phi}{\partial u^{b}}) + \frac{1}{2}
 \frac{\partial^2 \phi}{\partial s^2}
  + \frac{1}{2h} \frac{\partial h}{\partial s}
  \frac{\partial\phi}{\partial s}   - (\phi^2 - 1 ) \phi =0.
 \end{equation}
We see that the expansion parameter, i.e. the width $l_0 \equiv 1/M$,
 appears in front of the highest
order derivative with respect to $\tau$ present in this equation in such a
manner that in the zeroth order approximation the $\tau$ derivative drops out
(as well as in the first order approximation in this particular example).
It is well-known that in such cases straightforward perturbative expansion
in positive powers of the expansion parameter in general is not correct
 \cite{24}.

 The energy-momentum tensor in our model has the following components
 \begin{equation}
 T^{\mu\nu} = \partial^{\mu}\Phi \partial^{\nu}\Phi + \eta^{\mu\nu} L,
 \end{equation}
 with $L$ given by formula (1). The laboratory frame energy $E$ is given
  by the  integral
  \begin{equation}
  E=\int d^{3}x T^{00}.
  \end{equation}
  In this formula the field $\Phi$, which is a scalar with respect to
  coordinate transformations, can be regarded as a function of the
  $(u^a,\xi)$ variables. With the help of a formula for differentiation
  of composite functions  $T^{00}$ can be written in the
   following form
\begin{eqnarray}
\lefteqn{T^{00} = [(N^{-1})^{0a}\Phi_{,a} + n^{0} \Phi_{,\xi}]^{2} } \\
& &  +\frac{1}{2} g_{ab}(N^{-1})^{ac}(N^{-1})^{bd}\Phi_{,c}\Phi_{,d}
+\frac{1}{2} (\Phi_{,\xi})^2 + V(\Phi), \nonumber
\end{eqnarray}
where $a=0,1,2$; $n^{0}$ is the $\mu =0$ component of the
four-vector $(n^{\mu})$; the potential
\begin{equation}
V(\Phi) = \frac{\lambda}{2} (\Phi^2 - \Phi_{0}^{2})^2,
\end{equation}
and the matrix $N^{-1}$ is given by formula (15).

Also the volume element $d^{3}x$  is expressed by the co-moving coordinates,
\begin{equation}
d^{3}x =  \sqrt{-g} \frac{h(u^a,\xi)}{1-\xi K_{0a} g^{a0}}
d\xi d\sigma^1 d\sigma^2 .
\end{equation}

Observe that on the hyperplane of constant $x^{0}$ the $\tau$ variable
in general becomes a function of $\xi$; this is because
\begin{equation}
\tau + \xi n^{0}(\tau,\sigma^{i}) = x^{0}=constant.
\end{equation}
In general,  constant $x^{0}$ does not correspond to constant $\tau$,
 except for the membrane where $\xi=0$, hence $x^{0}=\tau$. This complicates
 very much calculation of the integral over $\xi$ because in general we do not
 know the explicit form of the $\tau$ dependence -- for this one would
 have to know explicit solutions of the Nambu-Goto equation for the membrane.

There is a particular case in which the calculation of the energy is
relatively simple: when each piece of the membrane associated with
 the domain wall is  at rest at the  initial time $x^0$.
Mathematically, this means that  the first derivative of $\vec{Y}$
 with respect to $x^0$ vanishes. Then, one can show that
 $n^{0}=0$, hence $x^0 = \tau$ for all $\xi$.

\section{ Solution by the Hilbert-Chapman-Enskog method}

Following the earlier papers we expand the field $\phi$ in
the non-negative powers of $1/M$,
\begin{equation}
\phi = \phi^{(0)} + \frac{1}{M} \phi^{(1)} +  \frac{1}{M^2} \phi^{(2)}
  + \frac{1}{M^3}  \phi^{(3)} + ... ,
  \end{equation}
  and we solve equation (18) perturbatively. The expansion parameter is $1/M$
and not $1/M^2$ because $1/M$ in the first power is present in the $h$ and
$G^{ab}$  functions after passing to the $s$ variable, formulae (17).

Inserting the perturbative Ansatz into equation (18) and expanding the l.h.s.
of it in powers of $1/M$ we obtain the following sequence of equations for
$\phi^{(n)}$:
\begin{equation}
\frac{1}{2}  \partial^2_s \phi^{(0)} +  \phi^{(0)} [1- (\phi^{(0)})^2] = 0,
\end{equation}
\begin{equation}
\frac{1}{2}  \partial^2_s \phi^{(1)}  +
 [ 1 - 3 (\phi^{(0)})^2 ] \phi^{(1)} =  K^a_a \partial_s \phi^{(0)},
 \end{equation}
\begin{eqnarray}
\lefteqn{\frac{1}{2} \partial^2_s \phi^{(2)}  +
 [ 1 - 3 (\phi^{(0)})^2 ] \phi^{(2)} =
 - 2  \Box^{(3)} \phi^{(0)}    }  \\
 & & +  2  K^b_a K^a_b \:s  \partial_s \phi^{(0)}
 + 3 \phi^{(0)} (\phi^{(1)})^2  +   K^a_a \partial_s \phi^{(1)}, \nonumber
 \end{eqnarray}
  \begin{eqnarray}
\lefteqn{\frac{1}{2} \partial^2_s \phi^{(3)}  +
  [ 1 - 3 (\phi^{(0)})^2 ] \phi^{(3)} =
4  K^a_c K^b_a K^c_b \: s^2 \partial_s \phi^{(0)}
 - 2  \Box^{(3)} \phi^{(1)} } \\
 & &
   + 2  K^b_a K^a_b \:s \partial_s \phi^{(1)}
 + 6 \phi^{(0)} \phi^{(1)} \phi^{(2)}  +  (\phi^{(1)})^3
   +   K^a_a \partial_s \phi^{(2)}.  \nonumber
 \end{eqnarray}
In these equations
\begin{equation}
\Box^{(3)} \equiv \frac{1}{\sqrt{-g}} \frac{\partial}{\partial u^{a}}
(\sqrt{-g} g^{ab}\frac{\partial}{\partial u^{b}})
\end{equation}
is the three-dimensional d'Alembertian on the
 world-volume $\tilde{\Sigma}$ of the membrane.
Notice that only Eq.(26) is nonlinear. The other equations are
 linear inhomogeneous  ordinary differential equations for
  $\phi^{(n)}, n\geq 1$.

In general, the range of the variable $s$ in these equations is restricted,
 $\frac{M}{2} \xi_{-} \leq s \leq \frac{M}{2} \xi_{+}$, because the
 coordinates $(u^a, \xi)$ do not cover the
 whole space-time. The values of $\xi_{\pm}$ are found from the condition
 that $h(u^a,\xi) \geq 0$ for all $\xi$ in the interval
 $(\xi_{+}, \xi_{-})$. In general $\xi_{\pm}$ depend on
   $u^a$. It is clear that the function $h$ with  generic $K_{ab}$ and
 $g_{ab}$ can vanish for finite $\xi$, see formula (12).
   From a mathematical point of view  there is no problem - just one
    has to add more coordinate maps,
 smoothly matched with the coordinates $(u^a, \xi)$, in
 order to cover the whole space-time. From a practical, calculational
 point of view this would be a significant complication with rather little
 gain in understanding the physics of the not-too-much-curved domain
 wall we consider in  this paper, because the additional
 coordinate maps would cover the
 uninteresting from physical point of view region in which the $\Phi$ field
 is exponentially close to one of the vacuum values. For this reason we do
 not introduce the additional coordinate maps, and therefore we do not
 determine the field in the region where  the domain wall exponentially
 merges into the vacuum.

Then another problem arises, namely how to impose the boundary conditions
for our solutions, because for finite $\xi = \xi_{\pm}$ the field $\Phi$
in general does not reach the vacuum values. This problem can be circumvented
in the following way. We shall find solutions of equations (26-29) assuming
that $s$ varies from $-\infty$ to $+\infty$. Notice that these equations
formally make sense for all $s$. The solutions have physical meaning
only in the restricted range of $s$.
Because we presume that the domain wall field for $\xi$ close to $\xi_{\pm}$
is exponentially close to the vacuum values, we require that for large $s$
$\phi^{(0)}$ is exponentially close to $\pm 1$, and $\phi^{(n)}, n \geq 1$,
 are exponentially close to zero. In this manner the boundary conditions at
  $\xi = \xi_{\pm}$ are fixed uniquely, eventhough in the somewhat implicit
  manner.

  The zeroth order equation (26) has the well-known solutions
\begin{equation}
\phi^{(0)}_{\pm}(s) = \pm tanh s.
\end{equation}
They obey the boundary conditions specified above.
We take $\phi^{(0)}_{+}$  as the
 starting point for the approximation procedure. In the case of
planar domain wall this $\phi^{(0)}$ coincides with the exact solution (3).
The corrections given by $\phi^{(n)}, n\geq 1,$ can be regarded as being due
to the curvature of the domain wall.

Let us pass to the first order equation (27). On the r.h.s. of it we have
the function
\begin{equation}
\partial_{s} \phi^{(0)}_{+} = \frac{1}{cosh^{2}s}.
\end{equation}
The crucial step in the Chapman-Hilbert procedure is to observe that the
operator
\[
\hat{L} \equiv  \frac{1}{2}  \partial^2_s +  1- 3 (\phi^{(0)})^2,
\]
present on the l.h.s. of Eqs.(27-29) has a zero eigenvalue. The coresponding
eigenfunction coincides with the function (32). (It is identical with
a zero mode for a kink in the 1+1 dimensional version of the considered
 model (1).) Multiplying the both sides of Eq.(27) by the function (32)
and integrating over $s$ in the interval $s\in(-\infty,+\infty)$ we find that
 the integral on the l.h.s. vanishes, while the r.h.s. is equal to
$\frac{4}{3}K^a_a$. Therefore, we have to accept the consistency condition
 for our expansion
 \begin{equation}
 K^a_a = 0.
 \end{equation}

It can  be shown that this condition is equivalent to Nambu-Goto equation
\begin{equation}
 \Box^{(3)} Y^{\mu} = 0
 \end{equation}
 for a relativistic membrane.  This is the reason for calling the surface
 $\tilde{S}$  the membrane. Solutions of the equation (33) for $Y^{i}$ have
 been discussed in numerous papers, see,e.g., \cite{16,23}.
 From Eq.(33) with fixed initial data one can calculate the $Y^i$ functions,
 and subsequently the extrinsic curvature coefficients $K_{ab}$ and the
 metric $g_{ab}$. Then, from the condition $h>0$ one can determine
  $\xi_{\pm}$.

Because of  condition (33) the r.h.s. of Eq.(27) in fact vanishes. Then, the
general vanishing at the infinity solution of Eq.(27) has the form
\begin{equation}
\phi^{(1)} = \frac{C(u^a)}{cosh^{2} s},
\end{equation}
where $C$ does not depend on $s$.

The next equation to be solved is Eq.(28). Because $\phi^{(0)}_{+}$ given by
formula (31) does not depend on $u^a$, and $K^a_a=0$,
there are simplifications on the r.h.s. of Eq.(28):
\begin{equation}
 \frac{1}{2}  \partial^2_s \phi^{(2)}  + [1 - 3 (\phi^{(0)})^2 ] \phi^{(2)} =
    2  K^b_a K^a_b  \frac{s}{cosh^{2}s}  + 3 C^2 \frac{sinh s}{cosh^{5}s}.
 \end{equation}
Again on the l.h.s. we have the operator $\hat{L}$, however this time the
integration with the zero mode (32)  gives the identity 0=0 because the r.h.s.
of Eq.(36) is an odd function of $s$.

 General solution of Eq.(36) can be found by a standard technique
 consisting of first finding two independent solutions of the homogeneous
 equation $\hat{L}\phi^{(2)}=0$, and next constructing a Green function,
see e.g. \cite{26}. As the two linearly independent solutions
 of the homogeneous counterpart of Eq.(36) we take
\begin{equation}
 \phi^{(2)}_{1} = \frac{1}{cosh^{2}s},
\end{equation}
  \begin{equation}
 \phi^{(2)}_{2} = \frac{1}{8} sinh(2 s) + \frac{3}{8} tanh s +
\frac{3}{8}  \frac{s}{cosh^{2}s}.
\end{equation}
As the Green function we take
\[
G(s,x) = 2 \phi^{(2)}_{2}(s)  \phi^{(2)}_{1}(x)  \Theta(s - x)
    - 2  \phi^{(2)}_{1}(s)   \phi^{(2)}_{2}(x)  [\Theta(s-x) - \Theta(-x)].
 \]
The general solution of Eq.(36) has the form
\begin{equation}
 \phi^{(2)}  = \alpha \phi^{(2)}_{1}   + \beta \phi^{(2)}_{2}
 + \int^{+\infty}_{-\infty} dx G(s, x) f(x),
\end{equation}
where $\alpha(u^a), \beta(u^a)$ do not depend on $s$, and
\begin{equation}
f \equiv 2  K^b_a K^a_b  \frac{s}{cosh^{2}s}
 + 3 C^2 \frac{sinh s}{cosh^{5}s}
\end{equation}
is the r.h.s. of Eq.(36). Formula (39) gives
\begin{equation}
 \phi^{(2)} = c_{1}(u^a,s) \phi^{(2)}_{1}(s) + c_{2}(u^a,s) \phi^{(2)}_{2}(s)
\end{equation}
where
\begin{equation}
c_{1}(u^a,s) =
 \alpha - 6 C^2 f_{1}(s) - 4 K^b_a K^a_b f_{2}(s),
\end{equation}
\begin{equation}
c_2(u^a,s) = \beta - \frac{C^2}{cosh^6 s} -
 4 K^b_a K^a_b f_{3}(s),
\end{equation}
The functions $f_{i}(s)$ are defined by the following formulae
\begin{equation}
  f_{1}(s) \equiv \int^{s}_{0} dx \frac{sinh x}{cosh^{5}x}
 \phi^{(2)}_{2}(x), \;
f_{2}(s) \equiv \int^{s}_{0} dx \frac{x}{cosh^{2}x} \phi^{(2)}_{2}(x),
\end{equation}
\begin{equation}
  f_{3}(s)\equiv - \int^{s}_{-\infty} dx \frac{x}{cosh^{4}x}.
\end{equation}
The functions $f_1, f_2$ are odd , while $f_3$ is even.
The function $\phi^{(2)}_{2}$ is given by formula (38). Because this function
exponentially grows for $s \rightarrow \pm \infty$, the coefficient function
$c_2$ has to vanish in this limit. Therefore, we have to
put
\begin{equation}
\beta = 0,
\end{equation}
while $\alpha$ is still arbitrary.

One more restriction comes from the third order equation (29).
Integrating it with the zero mode we obtain the following condition
\begin{equation}
 \Box^{(3)}C   +   (\frac{\pi^2}{4}-1)   K^{b}_{a} K^{a}_{b}\:  C
 + \frac{9}{35} \:  C^3 =  (\frac{\pi^2}{6}-1) \:
  K^{a}_{c} K^{b}_{a} K^{c}_{b}.
 \end{equation}
Notice that the function $\alpha$ is not present in (47) -- it drops out
because an integral giving the coefficient in front of it vanishes.
The coefficients in Eq.(47) has been obtained from  computations
of integrals over $s$.

Condition (47) restricts possible choices of the function
$C(u^a)$. It has the form of a non-linear 2+1 dimensional
wave equation. The function $C$ can be regarded as a scalar field
on  the Nambu-Goto manifold determined from equation (33).

Let us summarize results of our calculations. Up to the second order
in $1/M$
\begin{equation}
    \phi = tanh s + \frac{1}{M} \frac{C}{cosh^{2}s} +
 \frac{1}{M^2} \frac{\alpha}{cosh^{2}s} \;\;\;\;\;\;\;\;\;\;\;\;\;\;\;
 \end{equation}
 \[  \;\;\;\;\;\;
-  \frac{1}{M^2} [ C^2 (\frac{6 f_{1}(s)}{cosh^{2}s} +
  \frac{\phi^{(2)}_{2}(s)}
 {cosh^{6}s}) + 4  K^b_a K^a_b (\frac{f_{2}(s)}{cosh^{2}s} +
 f_{3}(s) \phi^{(2)}_{2}(s)) ],  \]
where $C$ obeys Eq.(47), while for $\alpha$ no explicit restrictions
have been obtained till now. However, such restrictions might appear as
the consistency condition following from the fourth order equation. Notice
that Eq.(47) for the coefficient $C$ introduced in the first order
solution (35) has been
obtained from the equation (29) two order higher. Thus, in order to claim that
formula (48) gives the domain wall solution up to the order $1/M^2$ one should
also find $\phi^{(3)}$ and check the consistency condition coming from the
fourth order equation.

The fourth order equation has the following form
 \[
  \frac{1}{2} \partial^2_s \phi^{(4)}  +  [ 1 - 3 (\phi^{(0)})^2 ]
 \phi^{(3)} =
- 2 \Box^{(3)} \phi^{(2)}
-  \frac{1}{\sqrt{-g}} \frac{\partial}{\partial u^{a}}
(\sqrt{-g} K^{ab}\frac{\partial C}{\partial u^{b}}) \frac{8 s}{cosh^{2}s}
\]
\begin{equation}
\;\;\;\;\; + 3 \phi^{(0)}_{+} (\phi^{(1)})^2 +
 6 \phi^{(0)}_{+} \phi^{(1)} \phi^{(3)} +
    3 (\phi^{(1)})^2  \phi^{(2)}
    \end{equation}
    \[   \;\;\;\;\; + 2  K^b_a K^a_b \: s \partial_s \phi^{(2)}
   +  4   (K^b_a K^a_b)^2 \: s^3   \partial_s \phi^{(0)}_{+}
+     4  K^a_c K^b_a K^c_b  \: s^2   \partial_s \phi^{(1)}.
 \]
The r.h.s. of this equation has already been simplified by taking into
account the condition (33) and the fact that $\phi^{(0)}_{+}$ does not depend
on $u^a$. $\phi^{(3)}$ is given by a formula analogous to (39).
Repeating once again the calculation leading to the
 consistency conditions we obtain from Eq.(49) the following
condition
\begin{equation}
 \Box^{(3)} \alpha   + \frac{1}{2}   K^{b}_{a} K^{a}_{b} \: \alpha
 -  \frac{36}{35}  C^2 \: \alpha = 0.
 \end{equation}
Thus, the fourth order equation (49) gives the restriction (50) for
$\alpha$.

The final step in our procedure is to make a convenient choice of the
initial position and velocities of points of the membrane. Its evolution
at later times is then determined by the Nambu-Goto equation (33). Let us
recall that the membrane is the auxilliary object merely, used
to define the co-moving coordinate system, so that the choice of the
initial data has no physical meaning.
The most natural choice is that at the initial value of $\tau$
 the position
of the membrane as well as its velocities coincide with those of the core.
By the definition of the membrane, for its points $s=0$. On the other hand,
we find from formula (48) that for the core (i.e. for the points where
$\phi$ vanishes)
\begin{equation}
s = s_0 \equiv - \frac{C}{M} - \frac{\alpha}{M^2} + {\cal O}(M^{-3}).
 \end{equation}
Therefore, our choice of the initial data for the membrane
 means that $s_0=0,\; \partial_{\tau}s_0=0$
at the initial value of $\tau = \tau_0$. This implies that
\begin{equation}
C(u^a, \tau_0) = 0,  \;\;
\frac{\partial C}{\partial\tau} (u^a,\tau_0) = 0,
 \end{equation}
 \begin{equation}
 \alpha(u^a, \tau) = 0, \;\;   \frac{\partial \alpha}{\partial\tau}
 (u^a,\tau_0) = 0.
 \end{equation}
 Notice that the more general solution $\alpha=-MC$ is not acceptable,
 because one should not mix different orders of the perturbative expansion.
 With the initial conditions (53) equation (50) implies that
 \begin{equation}
 \alpha = 0
 \end{equation}
 for all $\tau$. On the other hand, the term on the r.h.s. of equation (47)
 acts as an external, curvature dependent force which  in general
  leads to nontrivial $\tau$-dependence of the function  $C$
  even for the  trivial    initial data (52).

Thus, the domain wall solution constructed to the second order in $1/M$
with the Hilbert-Chapman-Enskog method has the form (48) with $\alpha=0$ and
$C$ determined from the equation (47) with the initial data (52).

The formula (51) gives the one-to-one mapping between the core and the
membrane: the point $(u^a, s=0)$ of the membrane is mapped to the point
$(u^a, s=s_0)$ of the core.

 \section{ Examples }

In this Section we specify our general solution to describe simple,
symmetrical domain walls.

(a)\underline{ The planar domain wall.}
Let us start from the simplest domain wall: the planar, static one. In our
formalism this means that the membrane is a motionless plane. In this case
$K_{ab}=0$, so that the external force in Eq.(47) vanishes. Therefore
$C=0$, and the solution (48) coincides with the exact solution (3).
Energy per unit square of the membrane is equal to
 \[  \frac{E}{|\tilde{S}|} = \frac{2}{3} M \Phi_{0}^{2}, \]
where $|\tilde{S}|$ denotes the area of a piece of the membrane.

(b)\underline{ The cylindrical domain wall.} Now the membrane is a cylinder
of the radius $r$. We assume that the cylinder does not move as a whole.
 Using the formulae of the Section 2 one can easily
calculate the extrinsic curvatures $K_{ab}$ and the metrics $g_{ab}$.
The Nambu-Goto equation has the following form
\begin{equation}
 \frac{\ddot{r}}{1-\dot{r}^2} + \frac{1}{r} = 0,
\end{equation}
where the dots denote derivatives with respect to $\tau$. This equation
has the "energy" first integral given by the formula
\[ E^{N-G} = \frac{r}{\sqrt{1-\dot{r}^2}},  \]
where $E^{N-G}$ is a constant. We also find that
\begin{equation}
 K^{a}_{b} K^{b}_{a} = \frac{2}{r^2(1-\dot{r}^2)}=
 \frac{2(E^{N-G})^2}{r^4},
  \;\;\;\;\;   K^{a}_{c} K^{b}_{a}  K^{c}_{b}  = 0.
  \end{equation}
 Thus, the external force in the equation (47) vanishes, and the membrane
 coincides with the core for all times --  the zeros of the
 scalar field follow the Nambu-Goto trajectory.

 The solution $\phi$ is given by formula (48) with $C=\alpha=0$
 and $K^a_bK^b_a$ as in the first of formulae (56).

The energy for the cylindrical domain wall solution is also easily calculated
 with the help of formulae (20),(21), (23). Because of the difficulty
 mentioned at the end of Section 2, we calculate the energy at the initial
 $\tau = \tau_0$, assuming that the core is then at instant rest. Thus,
 \[ r(\tau_0)=\dot{r}(\tau_0)=0, \]
 and $C$ has the initial value (52).
 At later time
 \[ r(\tau) = r_{0} cos\frac{\tau}{r_0},  \]
 as it follows from Eq.(55). It is easy to see that $E^{N-G} = r_0$.

 Let us notice here that our formulae are applicable only for
 $\tau/r_0 \in [0, \pi/2 - \delta]$, with some $\delta > 0$. The reason
 is that for $\tau/r_0$ approaching $\pi/2$ it no longer true that on the
 inner side of the domain wall the field $\Phi$ is  exponentially close to the
 vacuum. The value of $\delta$ can be estimated analogously as in the first
 of papers \cite{23}.

   The result for the energy per
 unit square of the membrane is the same as for the planar domain wall.
 Thus, the extrinsic curvature of the cylinder does not influence the total
 energy per unit square of the membrane (to the order $1/M^2$).
 However, the density of energy across the domain wall is changed in
 comparison with the planar case. Its maximum still lies on the core,
 but it is flattened: the maximal value of the energy density is equal to
 \[ \frac{M}{2} \Phi^2_0 (1 -\frac{8 f_{3}(0)}{M^2 r_0^2} ), \]
where $f_{3}(0) = \frac{2}{3} ln2 -\frac{1}{6} \approx 0.30,$
 and $r_0$ is the initial radius.

 (c)\underline{The spherical domain wall} The membrane is a sphere of the
 radius $r$. The Nambu-Goto equation has the following form
 \begin{equation}
  \frac{\ddot{r}}{1-\dot{r}^2} + \frac{2}{r} = 0.
  \end{equation}
Its solutions have been considered in, e.g., \cite{16,23}. $r(\tau)$ is given
 by an elliptic function and it is oscillating with $\tau$.
    The "energy" integral is given   by the formula
 \[ E^{N-G} = \frac{r^2}{\sqrt{1-\dot{r}^2}}, \]
 where $E^{N-G}$ is a constant.   Again we
 consider evolution of the domain wall only in a part of the first quarter
 of the cycle of the oscillations predicted by Eq.(57), before the spherical
 domain wall collapses to the center.

 For $K^a_b K^b_a$ we have the formula
 \begin{equation}
 K^{a}_{b} K^{b}_{a} = \frac{6}{r^2(1-\dot{r}^2)} =
  \frac{6 (E^{N-G})^2}{r^6}.
  \end{equation}
 The external force in Eq.(47) does not vanish:
\begin{equation}
   K^{a}_{c} K^{b}_{a}  K^{c}_{b}  = \frac{6}{r^3(1-\dot{r}^2)^{3/2}} =
     \frac{6 (E^{N-G})^3}{r^6}.
 \end{equation}
 Formulae (58), (59) have been obtained by taking $u^1 =\phi, u^2=\theta$,
 where $\phi, \theta$ are the usual azimuthal and polar angles
 parametrising the spherical membrane. The vector $\vec{n}$ has been
 directed to the outside of the sphere. Then $s>0$ ($s<0$) corresponds
 to the outside (inside) of the spherical membrane.

  Equation (47) has now the following form
  \begin{equation}
  \ddot{C} -  6(\frac{\pi^2}{4}-1)\:  \frac{C}{r^2}
   - \frac{9}{35} (1-\dot{r}^2) C^3 =
   -  \frac{\pi^2-6}{r^3 \sqrt{1-\dot{r}^2}}.
   \end{equation}
   At the initial instant $\tau=\tau_0$ : $C=0,\; r=r_0,\; \dot{r}=0,$
   and the acceleration $\ddot{C} = - (\pi^2-6)/r_{0}^{3} <0.$  Thus,
   for $\tau > 0$  $C$ will
   become more and more negative. It follows from formula (51) that for the
   core $s=s_0=-C/M >0.$ This means that the core lags behind the Nambu-Goto
   membrane.

 For the total energy per unit square of the membrane for the spherical
 domain wall we obtain a positive correction due to the
 curvature of the sphere
 \begin{equation}
 \frac{E}{4\pi r^2_0} = \frac{2}{3} M \Phi^2_{0}
 (1 + \frac{3 d_0}{M^2 r^2_0}),
 \end{equation}
 where
 \[ d_0 \equiv \int^{+\infty}_{-\infty} \frac{s^2}{cosh^{4}s} ds =
 \frac{\pi^2}{9} - \frac{2}{3} \approx 0.43.   \]
 As usual, we have calculated the energy at the initial instant;
then $r=r_0,\: \dot{r}=0;\; C=0,\: \partial_{\tau}C=0.$
  We see that the total energy $E$ differs  from the zeroth order result
  by the constant term  $8\pi d_0 \Phi^2_0/M$. The maximum of the
   energy density at the initial instant lies on the core, and
the maximal value of the energy density is equal to
     \[ \frac{M}{2} \Phi^2_0 (1 -\frac{24 f_{3}(0)}{M^2 r_0^2}). \]
  Thus, the energy density is  more flattened than in the case of
  cylindrical domain wall.

  The facts that the core shrinks more slowly than the Nambu-Goto membrane,
  and that the energy per unit square has the positive correction increasing
  with the curvature, mean that the spherical domain wall exhibits an
  effective rigidity.

\section{Remarks}

(a) In our opinion, the main advantage of the method proposed in this paper
is that it yields the domain wall solution without invoking the
effective action. In this manner we avoid many problems of the
effective action method, e.g. handling a model with higher derivatives.
Another advantage is that the co-moving coordinate system is tied
always to the Nambu-Goto type membrane. Equations of motion for this
type of membrane are probably the simplest ones to be dealt with.

In comparison with the polynomial approximation, \cite{23}, in which one
approximates the field inside the domain wall by a simple polynomial, the
present approach has the advantage that it is based on the expansion
with respect to the clearly defined parameter, i.e. the width. Therefore,
 it is easy to point out the limit in which the zeroth order
 solution is more and more accurate:  $C/M$ and
 $K^a_b K^b_a/M^2$ tending to zero. The polynomial solution never is exact.

(b) The corrections $\phi^{(1)}, \phi^{(2)}$ to the zeroth order
solution do not change significantly the picture  of the evolution of
the domain wall obtained from the zeroth order solution. For example,
it is easy to see that in the case of spherical domain wall the
difference between the radia of the core and of the membrane is of the
order $l^2_0/r_0$ -- it is smaller than the width of the domain wall.
Nevertheless, it is nice to know how to construct the more accurate
solution. Moreover, there are questions which one cannot answer without
detailed knowledge of the internal structure of the domain wall.
Let us mention here the problem of excitations of the domain wall. They
 can be regarded as bound states of certain particles with the
domain wall. In order to calculate spectrum of such excitations one
has to know details of the shape of the scalar field inside of
the domain wall.

(c) In the present paper we have completely neglected the possibility
that the field $\Phi$ can have a component oscillating with the
frequency of the order $M$. Because $M$ is the mass of the scalar
particle present in the non-topological sector of the model (1), one
should expect that such component is in general present. It could be
regarded as describing an emission or absorption of the scalar particles
by the domain wall. Such oscillating component can be calculated in
an extended perturbative expansion, which could be constructed, e.g.,
by generalising considerations of Section III of the first of papers
\cite{23}.

The problem of excitations of the domain wall and the problem of
radiation probably deserve a separate investigation.

\end{document}